# Effect of annealing on the charge-voltage characteristics of the SrBi$_2$(Ta$_x$Nb$_{1-x}$)$_2$O$_9$ films

N.V. Morozovsky [1], A.V. Semchenko [2], V.V. Sidsky [2], V.V. Kolos [3], A.S. Turtsevich [3], E.A. Eliseev [4], and A.N. Morozovska[1],[*]

[1] *Institute of Physics, National Academy of Sciences of Ukraine, 46, pr. Nauki, 03028 Kyiv, Ukraine*
[2] *F. Skorina Gomel State University, Sovetskaya 104, Gomel, 246019, Belarus;*
[3] *JSC «INTEGRAL», Kazintsa 121A, Minsk, 220108, Belarus;*
[4] *Institute for Problems of Materials Science, NAS of Ukraine, Krjijanovskogo 3, 03142 Kiev, Ukraine*

**Abstract**

The effect of changes of the Nb content and annealing on charge-voltage and current-voltage characteristics of film structures Pt/SrBi$_2$(Ta$_{1-x}$Nb$_x$)$_2$O$_9$/Pt/TiO$_2$/SiO$_2$/Si-substrate structures with x = 0, 0.1, 0.2-was studied theoretically and experimentally.

Theoretical modeling, which takes into account the mobile charged donors impact on the features of charge-voltage and current-voltage characteristics of ferroelectric-semiconductor films, **revealed** the changes of conductivity value and ferroelectric parameters. The results of theoretical analysis and experimental results are in qualitative agreement.

## 1. Introduction

Layered ferroelectic perovskites strontium bismuth tantalate, SrBi$_2$Ta$_2$O$_9$ (SBT), vanadate, SrBi$_2$V$_2$O$_9$ (SVN) and niobate, SrBi$_2$Nb$_2$O$_9$ (SBN) as well as their solid solutions attract permanent scientific interest due to their potential applications in non-volatile ferroelectric memories (NvFRAM) and their intriguing electronic, ferroelectric and electrophysical properties [1,2,3,4].

In particular, bismuth-layered ferroelectric compounds SBT and SBN (so-called Aurivillius phases with general chemical formulae Bi$_2$A$_{m-1}$B$_m$O$_{3m+3}$ [5] [6]) have been recognized as a prominent candidate for NvFRAM applications because of its negligible fatigue, low leakage currents, and ability to maintain ferroelectricity [7].

Intrinsic defects, chemical doping, grain orientation and post sintering annealing strongly influence microstructural, dielectric, piezoelectric, pyroelectric, ferroelectric and other electrophysical properties of SBT as for majority of perovskites. For instance, it was found by high-resolution *Z*-contrast imaging that defects in SBT single crystals can act as an efficient hole traps and can exert strong influence on its electronic and ferroelectric properties [7]. Vanadium addition of about 10% to SBT results in appreciable increase of dielectric constant and Curie temperature, while post sintering annealing leads to the increase of dielectric constant and to reduction of *dc* conductivity [8]. Microstructural, dielectric, pyroelectric and ferroelectric properties of the Aurivillius-type structure ceramics [5, 6] strongly depends on the grain orientation degree, direction relatively to pressing axis [5] and composition [6]. For the case of Sr$_{1-x}$Bi$_{2+x}$Ta$_{2-y}$Nb$_y$O$_9$ (SBTN) ceramics fabricated by a conventional technique Curie temperature and coercive field increase with Nb content *y* increase [9].

SBN ferroelectric powders, ceramics and thin films can be prepared by sol-gel method [10], single target RF-magnetron sputtering on Pt/Ti/SiO$_2$/Si substrates followed by rapid thermal process annealing or conventional furnace annealing [11] and other conventional ceramic fabrication techniques [9]. At that the thin film capacitors can be prepared with random and c-axis preferred orientation exhibiting excellent fatigue performances, remanent polarization of 12 μC/cm$^2$ and coercive field of 125 kV/cm [6, 11]. La$^{3+}$-doped SrBi$_2$Ta$_2$O$_9$ thin films for FRAM synthesized by sol-gel method [12].

---

[*] **corresponding author**

For the current work scope it is important to overview briefly some peculiarities of the fabrication technique of Nd-substituted $SrBi_2Ta_2O_9$ (SNBT) films and their physical properties. For the case of SNBT films sputtered on $Pt/Ta/SiO_2/Si$ substrates X-ray diffraction and photoelectron spectroscopy studies indicate that $Nd^{3+}$ is substituted preferentially at the $Sr^{2+}$ site into the layered SBT [13]. At that $Nd^{3+}$ substitution leads to increase of remanent polarization and coercive field.

Typically SNBT films reveal rather low fatigue (after $10^{10}$ switching cycles remanent polarization is decreased on 9% only) [13]. The reasons for the switching durability of SBT and SBN films and ceramics [1, 10, 11] were discussed in terms of the bulk ionic conductivities of the compounds [14]. Allowing for the fact that the bulk ionic conductivities of SBT and SBN are much higher than those of the perovskite ferroelectrics, authors conclude that their good switching durability is due to easy recovery of defects. Specifically, oxygen vacancies are easily released, resulting in limited space charge build up and domain wall pinning during the polarization reversal process [15]. At the same time it is known, that increase of annealing time in oxygen leads to decrease of oxygen vacancies concentration and other defects in the double oxide films [16, 17, 18]. The influence of oxygen losses during annealing on the parameters of the space charge in the depleted intergranular layers of $BaTiO_3$ and $SrTiO_3$ doped ceramics is well established [19,20].

Analysis of the abovementioned facts leads to the following conclusions:
1) There is a strong experimental evidence that pyroelectric, piezoelectric, ferroelectric and electro-physical properties of solid solutions based on $SrBi_2Ta_2O_9$, in particular for the solid solutions $Sr_{1-x}Bi_{2+x}Ta_{2-y}Nb_yO_9$ are controlled by doping and thermal treatment.
2) By adding 10-30 wt. % Nb to $SrBi_2Ta_2O_9$ one obtains solid solutions $Sr_{1-x}Bi_{2+x}Ta_{2-y}Nb_yO_9$ (SBTN) with high Curie temperature, spontaneous polarization and coercive field. Annealing procedure affects the properties of solid solutions in different ways, which requires additional experimental studies. There is the reason to assume a significant role of mobile defects, especially oxygen vacancies, for the properties of the films important for applications.
3) It is known that different annealing time of some double oxides films corresponds to a different concentration of oxygen vacancies acting as mobile donors. Also the effect of oxygen losses on the dielectric properties of some ternary oxides is known. However, there is no certainty that this is correct for Aurivillius family of layered oxides, in particular for films of SBTN.

Up to now, there is no theory, which could be compared with the experiment, explaining the influence of annealing and concentration of oxygen vacancies on the characteristics of the polarization reversal in films SBTN.

Based on the motivation 1) - 3), the aim of our work is theoretical and experimental study of the effect of annealing on the charge-voltage and current-voltage characteristics of the films of SBTN solid solution.

The paper is organized as follows. Description of the model and the main results of theoretical modeling of the charge-voltage and current-voltage characteristics of the films are given in Section 2. Experimental results for non-annealed and annealed films are given in sections 3 and 4, respectively. Section 5 contains the discussion.

## 2. Modeling of charge-voltage and current-voltage characteristics of the films

We considered ferroelectric semiconductor film with movable charged donors (positively charged oxygen vacancies, $V_O^{2+}$) and stationary niobium ions, $Nb^{3+}$ (an additional source of free electrons) as an approximation of the experimental structure. Thus mobile charge carriers are electrons and oxygen vacancies (donors). Schematic representation of the considered film with electrodes is shown in **Figure 1**.

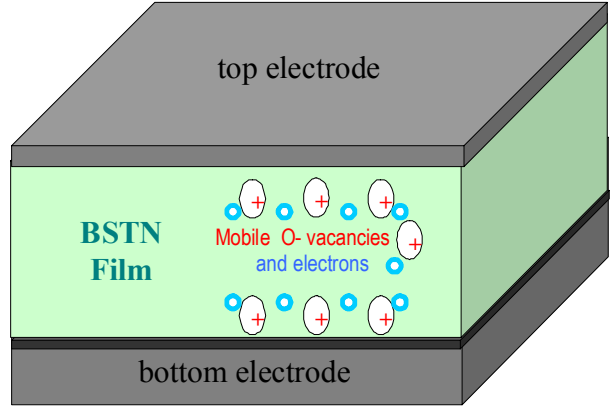

**Figure 1.** Schematic representation of film with mobile charge carriers and electrodes.

Theoretical modeling of the charge-voltage and current-voltage characteristics in ferroelectric semiconductors with mobile charged donors was performed in the framework of the continuum phenomenological theory of Ginzburg - Landau - Devonshire [21].

We numerically solved the system of coupled equations including the Ginzburg - Landau - Devonshire - Khalatnikov equations for the spatial distribution and evolution of polarization in the film, the continuity equations for electron and mobile donor currents, and the Poisson equation for finding the potential distribution depending on free charge density. The boundary conditions correspond to the ideal ohmic contact with electrodes, which were considered as completely transparent for electrons and blocking for $V_O^{2+}$. It was supposed that harmonic voltage, $U(t) = U\sin(\omega t)$, is applied to one of the electrodes, and another one is grounded.

The system of equations in dimensionless variables with the appropriate boundary conditions and typical ranges of material parameters are presented in papers [22, 23] together with the analysis of solutions of a system. Here we consider the transformation of the solutions under the temperature variation and the mobile defect concentration change.

Charge-voltage (**Fig. 2a**) and current-voltage (**Fig. 2b**) loops calculated for semiconductor-ferroelectric at a fixed concentration of mobile donors ($n_1$) and different temperatures $T_1 < T_2 < T_3$ are shown in **Figure 2**. Temperature values $T_{1,2,3}$ correspond to shallow ferroelectric phase, where the temperature dependence of the polarization is far from saturation. Note that the transition temperature for thin films can differ strongly from the Curie temperature of the bulk material due to size effects [24, 25].

Charge-voltage loop corresponding to $T_1$ (**Fig. 2a**) has a relatively saturated shape and characterized by a small slope at the coercive voltage. (Coercive voltage is defined as the voltage at which the polarization vanishes [26]).

Current-voltage loop for temperature $T_1$ has a relatively sharp peak above the coercive voltage. With increasing temperature to $T_2$ and $T_3$ the shape of charge-voltage loops (**Fig. 2a**) becomes unsaturated, slope of the loop increases, the coercive voltage is reduced and the value of remanent and maximal polarization decreases remarkably. At that the current maximum of the current-voltage loop (**Fig. 2b**) decreases, becomes smoother and shifts to lower voltages at the transition from $T_2$ to $T_3$. Such transformation of the loops is characteristic for ordinary ferroelectrics under approaching the ferroelectric - paraelectric transition from the low temperatures. Note that the shape of loops corresponding to $T_3$, is characteristic to the ionics, where no spontaneous polarization and so no ferroelectricity is observed [23] except the ferroelectric-ionics [33].

Performed analysis [23, 24] showed that the features of current-voltage characteristics (in particular position and shape of the maxima) are dependent on the concentration of defects. Therefore, it is worth to demonstrate how the shape of the charge-voltage and current-voltage loops change**s** with increasing concentration of mobile donors at fixed temperature.

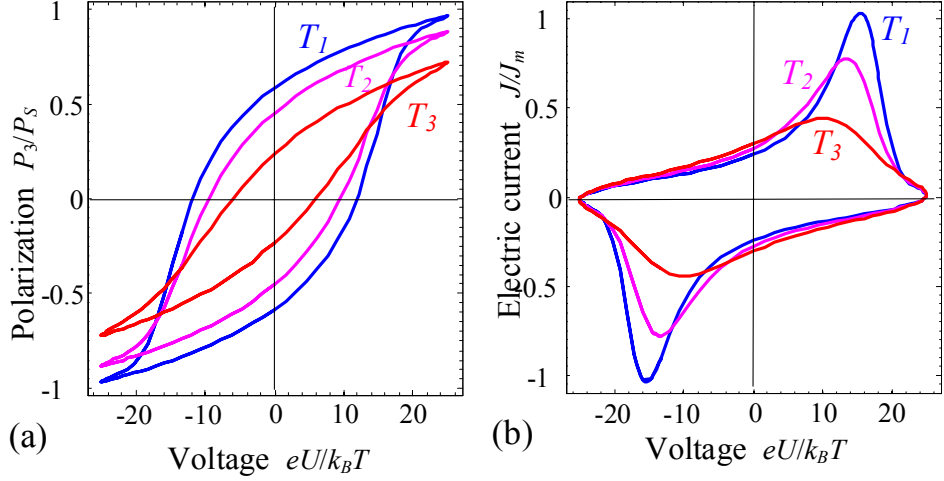

**Figure 2.** Charge-voltage (a) and current-voltage (b) loops for ferroelectric semiconductor at fixed concentration of donors and different temperatures $T_1 < T_2 < T_3$. Polarization and electric current are normalized to their maximal values. Voltage is given in $k_BT/e$ units at room temperature. Quantitative characteristics: the ratio of the maximum and remanent polarization $P_r/P_m$ = 0.63 for $T_1$, 0.51 for $T_2$, 0.32 for $T_3$ (**Fig. 2a**); and the ratio between maximal and zero-voltage currents $I_0/I_m$ = 0.3 for $T_1$, 0.36 for $T_2$, 0.73 for $T_3$ (**Fig. 2b**).

**Figure 3** shows the charge-voltage (**Fig. 3a**) and current-voltage (**Fig. 3b**) loops computed for doped semiconductor ferroelectric at a fixed temperature $T_1$ and concentrations of mobile donors $n_1 < n_2 < n_3$ which differs in one order of value.

Charge-voltage loop corresponding to the concentration $n_1$ is unsaturated (**Fig. 3a**), relatively narrow, with small remanent polarization and characterized by certain slope and relatively small coercive voltage. With increasing the concentration of mobile defects to $n_2$ the charge-voltage loop becomes less saturated and bends toward the voltage axis, the magnitude of the remanent polarization increases slightly, and the coercive voltage increases noticeably. It should be noted that the shape of the loop corresponding to $n_2$ is characteristic to nonpolar ionics, where the ferroelectric properties are not observed [23].

Charge-voltage loop for $n_3$ has a quasi-elliptical shape, without the typical for ferroelectric phase tendency toward saturation. The magnitude of the remanent polarization is reduced slightly. Such loops can be observed in the paraelectric phase of ferroelectrics when their permittivity and resistivity depend on the applied voltage.

The shape of current-voltage loop is also undergoing significant changes under the transition from $n_1$ to $n_3$ (**Fig. 3b**). Current-voltage loop computed for $n_1$ has maxima at voltages somewhat larger than the coercive voltage of corresponding charge-voltage loop. Current-voltage loop for $n_2$ has the shape of a smoothed parallelogram, and its diffuse maxima are spaced appreciably wider than the coercive voltage**s** of the charge-voltage loop. Current-voltage loop for $n_3$ has current maxima spaced much wider than the coercive voltage of corresponding to the charge-voltage loop (compare **Figs 3a and 3b**).

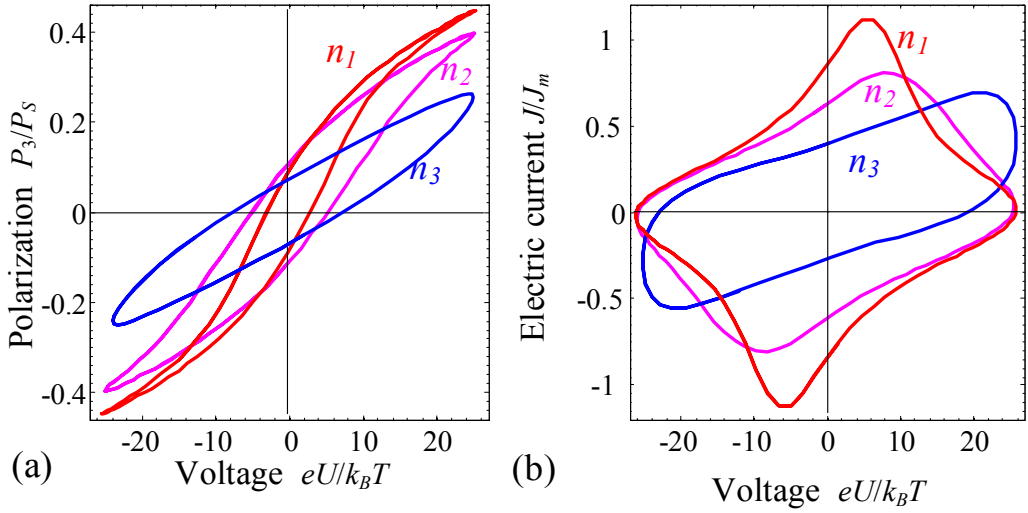

**Figure 3**. Charge-voltage (a) and current-voltage (b) loops for doped ferroelectric semiconductor at fixed temperature and different concentrations of mobile donors $n_1 < n_2 < n_3$. Polarization and electric current are normalized to their maximal values. Voltage is given in $k_BT/e$ units at room temperature. Quantitative characteristics: the ratio of the maximum and remanent polarization $P_r/P_m = 0.20$ for $n_1$, 0.27 for $n_2$, 0.25 for $n_3$, (**Fig. 3a**); and the ratio between maximal and zero-voltage currents $I_0/I_m = 0.77$ for $n_1$, 0.79 for $n_2$, 0.58 for $n_3$ (**Fig. 3b**).

Thus, theoretical modeling has shown that the increase of mobile donors concentration in films of ferroelectric semiconductor leads to the increase of coercive voltage and conductivity, which is generally expected and trivial result. However, a comparative analysis of changes of the loops shape caused by temperature variation at fixed concentration of donors and by varying the concentration of donors at fixed temperature showed that these changes is not trivial.

Indeed, under increasing both temperature or mobile donor concentration, similar is only decreasing of the slope of charge-voltage loops, but the position of current maxima changes in different way (compare **Figs 2a,b** and **3a,b**). Under transition from $T_1$ to $T_3$ the considerable decrease of remanent polarization, coercive voltages and current maxima without qualitative change of charge-voltage and current-voltage loops shape take place (see. **Fig. 3a**). Under transition from $n_1$ to $n_3$ moderate changing of remanent polarization and increase of coercive voltages is followed by qualitative change of loops shape (see. **Fig. 3b**). This is not consistent with the notion that the impact of charge carriers could shift the effective Curie temperature only [27], inasmuch as in this case the spontaneous polarization and coercive field should decrease and the shape of the loop should not change qualitatively.

## 3. Experimental

### 3.1. Preparation of SBTN films

The starting product for sol preparation and subsequent formation of the capacitor $SrBi_2Ta_{1-x}Nb_xO_9$ layer was 0.01 mol/l solution of inorganic metal salts (niobium chloride, bismuth and strontium nitrates in toluene).

Thermal $SiO_2$ layer of 300 nm - thickness was formed on Si (100) substrates of 0.675 mm - thickness, then the layer of 700 nm - thick boron phosphorus silicate glass (**BPSG**) with about 5.3 wt.% of B and 3.3 wt.% of P was deposited, which is subjected to burning-off at 850 ºC for 30 minutes in $O_2$ atmosphere.

Adhesion $TiO_2$ precoat was formed by magnetron sputtering of Ti film of 30 nm – thickness. Then the structure was annealed at 750 ºC for 60 min in $O_2$ atmosphere. Pt-electrode of 210 nm - thickness was deposited by magnetron sputtering, and subjected to crystallization annealing at temperatures from 800 ºC to 820 ºC for 30 min in $O_2$ atmosphere. Adhesive annealing of Pt was carried out at 400 ºC for 20 min in $N_2/H_2$-atmosfere immediately before deposition of the SBTN film.

Deposition of SBT and SBTN films on Pt/TiO$_2$/BPSG/SiO$_2$/Si-substrate was performed by spin-coating of stable solution at the substrate rotation frequency from 800 to 1500 rot/min during 2 to 5 seconds. Organic solvent was removed by multi-stage drying under temperature increase from 80 ºC to 350 ºC for 6 min.

Required **thickness** of SBT and SBTN films (200-300 nm) **was** obtained by layer-by-layer deposition of 2-3 sol-gel layers with heat treatment of each layer at temperature of 700 ºC. Perovskite SBTN structure was formed during annealing at (800-1000) ºC for 30 min.

"Top" Pt-electrodes of 70 nm - thickness and 2 mm diameter were deposited by magnetron sputtering through a shadow (aperture) mask.

### 3.2. Attestation of SBTN films

X-ray phase analysis was performed by ARL X'tra diffractometer (Thermo Ficher Scientific, Switzerland), using Bragg-Brentano geometry and reflected Cu K$\alpha$1-and K$\alpha$1-radiation. To reduce reflections from the substrate small incidence angles was used. Angular 2θ-range was selected from 10º to 60º.

Diffraction peaks identification was performed using database JCPDS (with software Search Match). Diffractograms were processed using JANA 2006 software. X-ray diffraction profiles obtained for SBT and SrBi$_2$Ta$_{2-y}$Nb$_y$O$_9$ with the ratio Nb/Ta y=10 wt. % (**SBTN10**) and y=30 wt. % (**SBTN30**) are shown in **Fig. 4**.

Comparison of the intensity and peak positions of crystallographic planes (115), (200) and (0010) showed that the SBTN10 composition after annealing at temperature of 800 ºC has the structure nearest to perovskite type. These qualitative estimations were confirmed by quantitative analysis of the phase composition performed using the JANA 2006 software (**Table 1**).

The results of scanning force microscopy (with SOLVER 47-PRO processed by the Gwiddion software) are presented in **Fig. 5**. For all the films the grain effect is visible. The increase of Nb content leads to change of the shape and size of the grains and their size dispersion. At that average grain size increases from 50 nm for SBT to 100 nm for SBTN10 with rounded off grains and to 300x700 nm for SBTN30 with preferable orientation of oblong grains (see **Fig. 5a,b,c**).

Results of scanning electron micrography of chips (cross section) and the surface morphology of annealed two-layer films of SBTN are presented in **Fig. 6**. Layered structure **(Fig. 6a, c)** corresponds to the sequence (SBT or SBTN)/Pt/TiO$_2$/SiO$_2$ at the thicknesses in the range (220 - 230) nm for SBT films and (260 - 270) nm for SBTN films. SBT film**s (Fig. 6a, b)** consist of partially fused granules with sizes ranging from 70 to 200 nm separated by pores with sizes from 20 to 50 nm. SBTN films **(Fig. 6c, d)** consist substantially of fused grains with sizes from 300 to 500 nm.

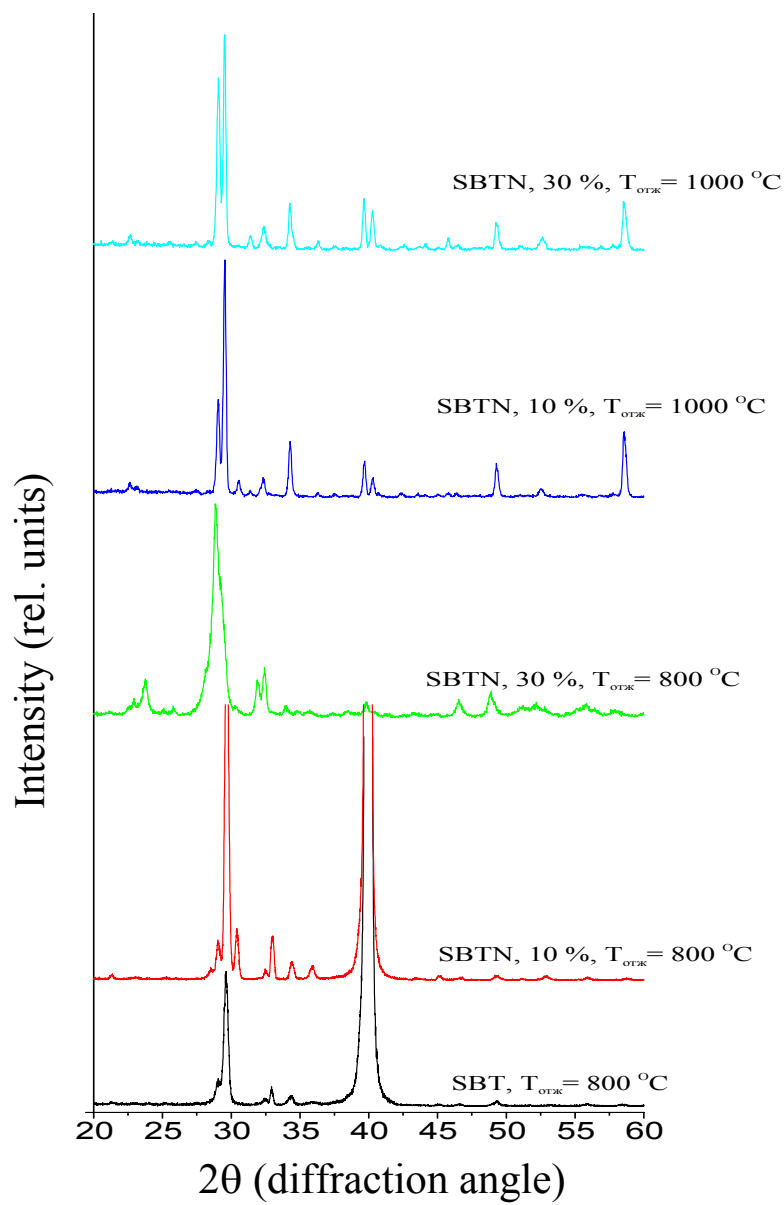

**Figure 4**. XRD patterns of SBTN films annealed at various temperatures.

**Table 1.** Results of quantitative analysis of the phase composition of SBTN films with different contents of Nb, prepared at different annealing temperatures.

| Material | Content of Nb, % | Annealing temperature, °C | Content of the perovskite phase, % |
|---|---|---|---|
| SBT | 0 | 800 | 65.2 |
| SBTN | 10 | 800 | 85.2 |
| SBTN | 30 | 800 | 77.3 |
| SBTN | 10 | 1000 | 67.2 |
| SBTN | 30 | 1000 | 59.3 |

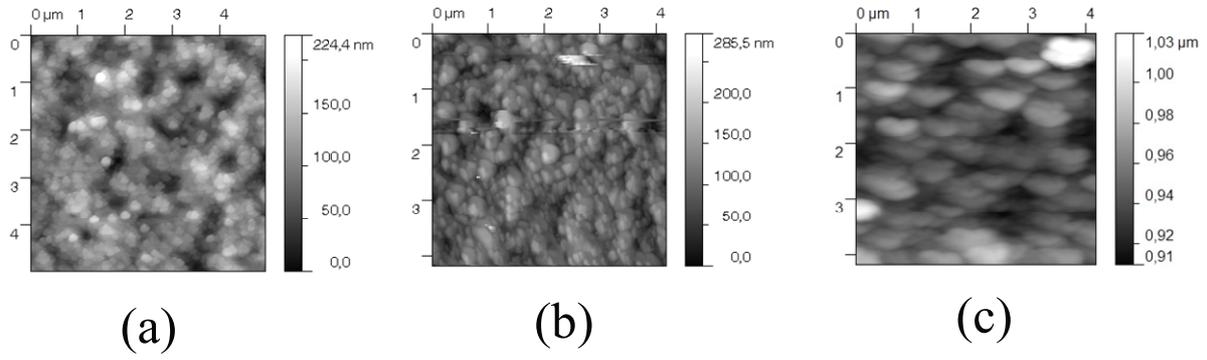

**Figure 5.** AFM images of different films: (a) – SBT, (b) – SBTN with 10% of Nb content and (c) – SBTN with 30% of Nb content.

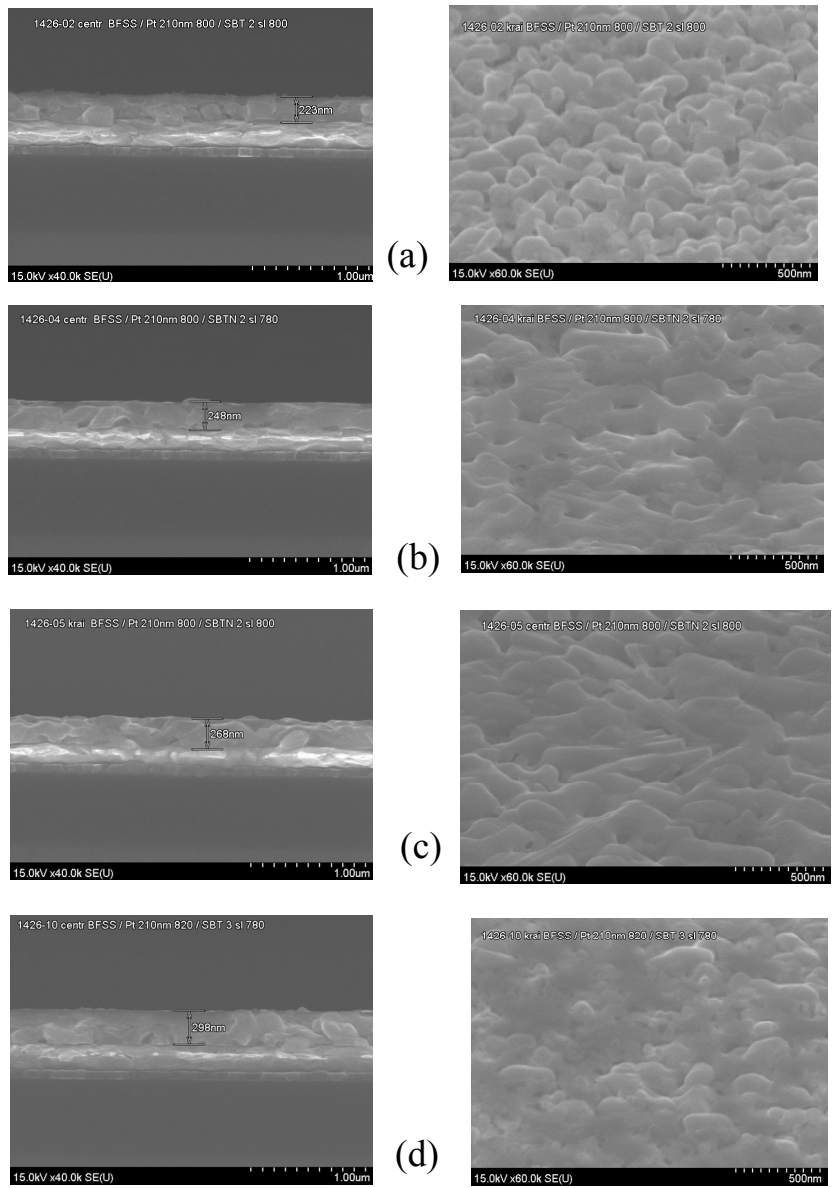

**Figure 6.** SEM images of chips (left side) and surfaces (right side) of two-layer SBT (a, d) and SBTN (b, c) films annealed at 800 ºC (a, c) and 780 ºC (b, d). Bottom Pt electrodes were annealed at 800 ºC (a, b, c) and 780 ºC (d). The sequences (SBT or SBTN)/Pt/TiO$_2$/BPSG/SiO$_2$ are well distinguished.

### 3.3. Resources and facilities of electrical measurements

Charge-voltage and current-voltage hysteresis loop**s** of Pt/SBTN/Pt/TiO$_2$/SiO$_2$/Si structures with different Nb content were obtained using a modified Sawyer-Tower circuit with C1-89 dual-beam oscilloscope. Charge-voltage loops of structures Pt/SBTN/Pt/TiO$_2$/SiO$_2$/Si at different annealing conditions were obtained using precision analyzer of semiconductors parameters HP4156V and probe station Micromanipulator-7000 and control computer with "EasyEXPERT" software. Measurements were carried out under the following ambient conditions: relative humidity 70%, temperature 22 °C and atmospheric pressure 750 torr.

### 4. Results of experiment

#### 4.1. Charge-voltage and current-voltage loops of non-annealed films

Dynamic charge-voltage and current-voltage loops for unannealed films of SBT and SBTN with different contents of Nb are shown in **Figure 7**. Shape of charge-voltage loops (**Fig. 7, top row**) is characteristic of the nonlinear dielectric with losses [26]. Form of current-voltage loops (**Fig. 7, bottom row**) reflects the rate of change of the charge, maximal in the vicinity of maxima. The values of the coercive voltage for charge-voltage loops are in the range (12-16) V (600-800 kV/cm) at 50 Hz, 50 V sine wave. This well exceeds the known values of 1 V [1] and 50 kV/cm [13] for SBT films and SBN/BTN ceramics [6].

Estimation of SBT and SBTN relative permittivity from measurements of the capacitance (at 1 kHz) of non-annealed films gives a value of ≈ 15, which is almost an order of magnitude less than the known value for SBT ceramics ≈ 130 [8]. This allows to consider the possible existence of local ferroelectric regions in non-annealed SBTN films. At that only a small fraction of the applied voltage drops on ferroelectrically active inclusions, which leads to the increase of apparent coercive voltage.

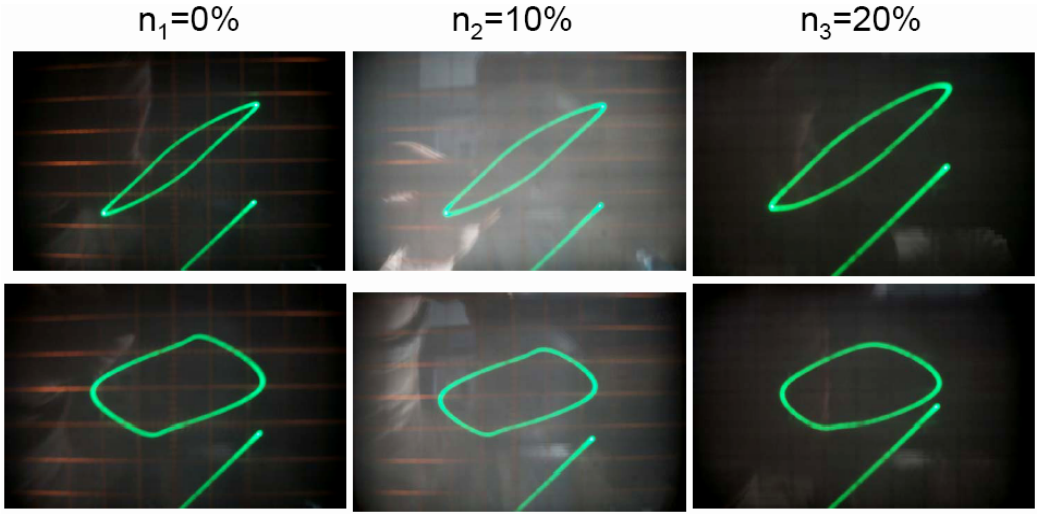

**Figure 7.** Dynamic charge-voltage (top row) and current-voltage (bottom row) loops for non-annealed SBT and SBTN films with different content of niobium ($n_1$ = 0%, $n_2$ = 10%, $n_3$ = 20%). The scale factors are 20 V/div. for horizontal axis, and 0.1 mC/div. of charge and 50 μA/div. for the current (vertically axes).

Comparison of the results of the experiment (**Fig. 7**) and theoretical modeling (**Fig. 3**) shows a qualitative agreement of the loops form and the character of their changes. For charge-voltage loops with Nb content increasing the decrease of saturation degree and increase of coercive voltage is observed (see **Fig. 7,** top row). For current-voltage loops current maxima are spaced along the voltage axis wider than the corresponding coercive voltages of charge-voltage loops (see **Fig. 7,** bottom row). With Nb content increase the distance between these maxima increases. These facts are in agreement with the computed behavior of charge-voltage and current-voltage loops with increasing concentration of mobile donors (see **Fig. 3**).

Small changes in the values of maximal and "residual" charge for charge-voltage loops and maximal current of current-voltage loops with increasing Nb content is qualitatively consistent with the theoretical analysis. As for the difference between trends of these changes, it may be related to the changes of the dielectric parameters and spatial characteristics of ferroelectric inclusions when changing the content of Nb unaccounted for in the theoretical model and existing in reality.

### 4.2. Dynamic charge-voltage loops of annealed films

The crystallization annealing (ca) of SBTN films was carried out at temperatures $T_{ca}$ from 750 °C to 770 °C for a time $t_{ca}$ from 30 to 60 min. Ferroelectric type features appeared on the loops after crystallization at $T_{ca}$ = 750 °C when remanent polarization $P_r$ about 2.1 µC/cm$^2$. At $T_{ca}$ = 770 °C $P_r$ value increased to 3.1 µC/cm$^2$ and at $T_{ca}$ = 800 °C reached 3.7 µC/cm$^2$. The increase of $t_{ca}$ from 30 to 60 min at $T_{ca}$ = 770 °C has a little effect on the coercive voltage and $P_r$ values.

Typical charge-voltage loops obtained for Pt/SBTN/Pt/TiO$_2$/SiO$_2$/Si structures at different annealing conditions are shown in **Fig. 8**. After formation of the upper Pt-electrode unstable asymmetric loops with rounded ends, low degree of saturation and small $P_r$ value were observed (**Fig. 8a**).

The shape of this loop is similar to that observed by authors of Ref. [1] after degradation caused by deposition of the SiO$_2$ layer on top of the SBT capacitor with Pt-electrodes and subsequent formation of contact windows by reactive ion etching. The recovery of symmetrical loops with high value of remanent polarization (about 5 µC/cm$^2$) was achieved by the annealing in oxygen atmosphere.

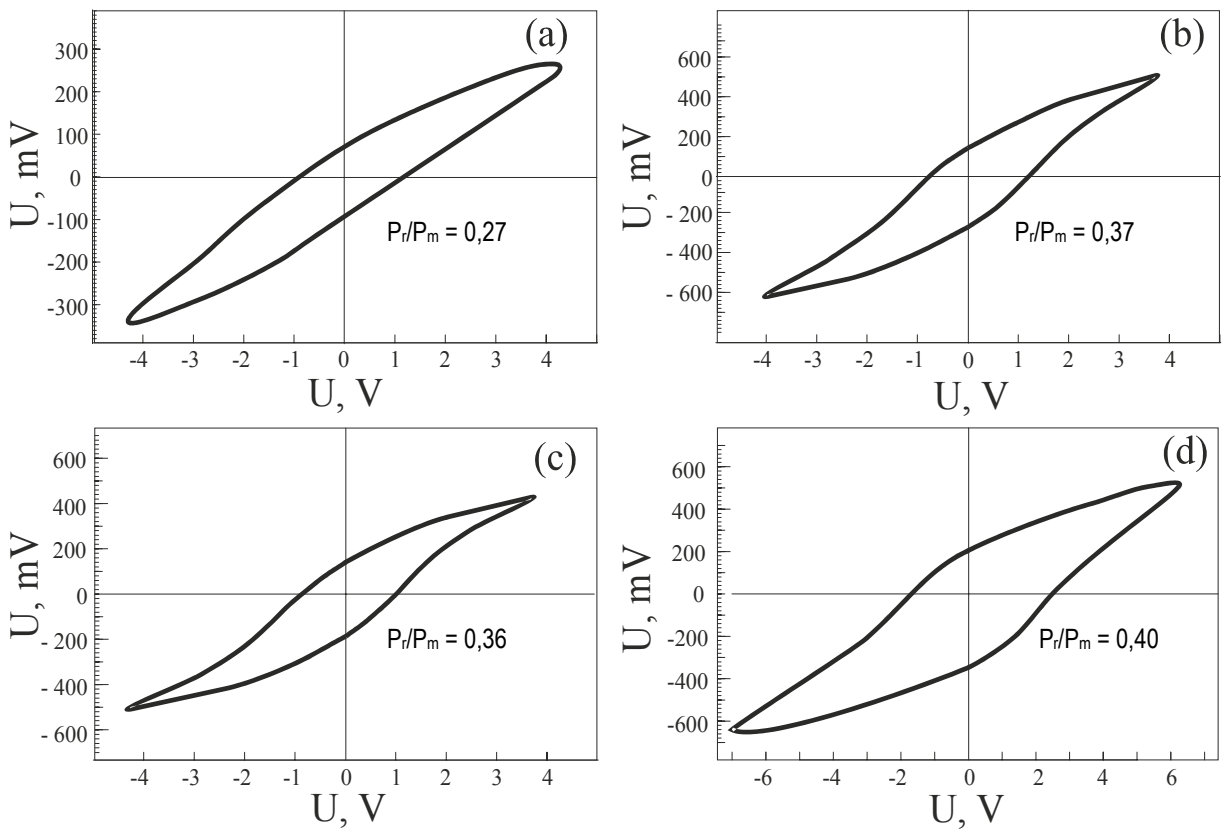

**Figure 8.** Typical hysteresis loops of structures with SBTN film, formed by annealing the Pt-electrode and crystallization annealing at different conditions: **(a)** 750°C after forming the upper Pt electrode before its annealing; **(b)** 750°C, 30 min. and 780°C; **(c)** 750°C, 60 min. and 780°C; **(d)** 780°C, 30 min. and 800°C. Note that the amplitude of the applied voltage in the part (d) was increased from 4 to 6.5 V due to the increase of the coercive voltage.

For studied SBTN-based structures the shape of loops approaches the ferroelectric one after low-temperature annealing in an oxygen atmosphere at 400 °C, exceeding the Curie temperature for SBT 305 C [8]. At that the refractive index of the SBTN film decreased from 2.1 to a typical value of 1.9 [?], which may indicate oxygen depletion after the formation of the upper electrode and the restoration of oxygen content after annealing.

The value of the coercive voltage is about 1 V (corresponding coercive field is 50 kV/cm) at $T_{ca}$ = 750 °C weakly depends on $t_{ca}$ (**Fig. 8a-c**), and reaches 2 V for $T_{ca}$ = 780 °C, $t_{ca}$ = 30 min. (**Fig. 8d**). It should be noted the proximity of the coercive voltage 1V to the data [1], and coercive field ≈ 50 kV/cm to the data [13], obtained for SBT films with thickness values 200 nm [1] and 220 nm [13]), and the proximity of the coercive voltage 2 V (corresponding coercive field is 100 kV/cm) to the data obtained for ceramics $(SrBi_2Nb_2O_9)_{0.35}(Bi_3TiNbO_9)_{0.65}$ [6].

## 5. Discussion

### 5.1. Shift of charge-voltage loop

A characteristic feature of the loops in **Fig. 8** is the presence of the horizontal (along the voltage axis) and vertical (along the charge axis) shifts. Horizontal shift is usually associated with the presence of internal bias field generated by a system of polar defects and causing a preferential orientation of the polarization [28].

The vertical shift of the loops is negative and depends on the annealing conditions. It is smallest for initial structures (-15 mV at U = 0 and -30 mV at maximum applied voltage) (**Fig. 8a**) increases to -70 mV and -50 mV for $T_{ко}$ = 750 C, $t_{ко}$ = 30 min (**Fig. 8b**) and decreases to -50 mV and -40 mV for $T_{ко}$ = 750 °C, $t_{ко}$ = 60 min. (**Fig. 8c**).

Vertical shift, which is like to the one observed in PZT films [29], can be attributed to the effect of rectification, which is a consequence of the asymmetry of RC nonlinearity of sub-electrode layers and the presence of ripple-through conduction determining the leakage current. The consequence of this asymmetry is the disparity of the electric charge transferred during positive and negative half-cycle of the alternating voltage. This leads to the reference capacitor charging that is connected in series with the sample; hence appeared DC voltage on the capacitor determines the vertical shift of the loop.

A circuit simulation using a linear series-parallel RC-circuit (**Fig. 9**, left), equivalent to the two-component multilayer structure [8, 30, 31], does not give a loop with vertical displacement. Introducing of diode type (D) series-opposing nonlinear elements, as proposed in the paper [32] and used for PZT films [29], (**Fig. 9**, right) allowed to obtain a shift of different sign and value by corresponding adjustment in both RCD-groups. Elimination of the resistance $R_b$ as an element equivalent to the leakage resistance leads to the disappearance of loop displacement.

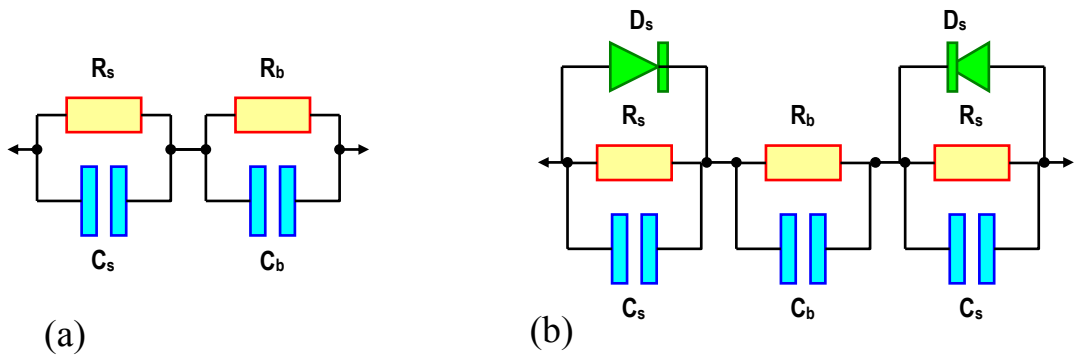

Figure 9. Equivalent circuit models: series-parallel linear RC (a) and nonlinear combined RCD (b).

### 5.2. Effect of annealing temperature on the parameters of the charge-voltage loops

Relatively small increase of $T_{ca}$ from 750 °C to 780 °C leads to changes in loop shapes, namely pointed ends of the loop (see **Fig. 8b**) changed to the smoothed ones (see **Fig. 8d**) and the breaks emerged in the vicinity of the coercive voltage (see Fig. 8d) instead of smooth curves (see Fig. 8b). This is an indication of the inhomogeneities appearance involved in the process of the polarization reversal due to the pinning phenomenon [33].

Self-consistent consideration [34] showed the presence of at least two contributions to ferroelectricity for monocrystalline SBT: the direct contribution through the O-Ta-O bonds in the perovskite groups $\{SrTa_2O_7\}^{2+}$ and the indirect contributions from the displacement of weakly bounded Bi in $\{Bi_2O_2\}^{2-}$ layers.

In the paper [7] it was shown that the formation the "planar" defects in the places of discontinuity! of Bi-O layers is inherent to SBT, at that the distance between adjacent perovskite Sr-Ta-O blocks in defective areas is smaller on the 1,2 Å than in the defect-free ones. This implies the presence of the pseudo-perovskite blocks distortions on the borders between defective and defect-free zones in the vicinity of these defects. This feature is clearly associated with the state of oxygen subsystem, and hence depending on the annealing conditions, it can simultaneously affect both the value of the polarization, and the conditions of the pinning.

Noting the role of $Bi^{3+}$ and $Ta^{5+}$ ions in capturing electron charge carriers [7, 34] one can assume that it is possible to bind defect of oxygen subsystem within the "planar" defects in SBT, like within the grain boundaries in the perovskite ceramic [19].

At elevated temperatures the weakly bounded $Bi^{3+}$ can leave the $\{Bi_2O_2\}$ layers and replace $Sr^{2+}$ in the adjacent areas to form not only Sr vacancies ($V_{Sr}^{2-}$) in A-positions of perovskite-like groups $\{SrTa_2O_7\}$ in the form $Bi_{2x}(V_{Sr})_xSr_{1-3x}$ [13] to keep charge neutrality, but also to release the excess oxygen in the places of discontinuity of layers $\{Bi_2O_2\}$ in accordance with the reaction

$$Bi_2O_2 \rightarrow 2Bi_{Sr} + 2O_O^{2-} + V_{Sr}^{2-}, \qquad (1)$$

where $Bi_{Sr}$ means $Bi^{3+}$ in the position of $Sr^{2+}$ (A-position), $O_O^{2-}$ means $O^{2-}$ in the position of oxygen. Therefore, the essential role of $V_{Sr}$ in the formation of polar properties of SBT [7, 13] can be supplemented with the role of oxygen defects consideration.

It is known that the doping by donors can facilitate the losses of oxygen during annealing of ceramics $SrTiO_3$ [30, 20] in accordance with the equilibration reaction in the following form:

$$2O_O \leftrightarrow O_2(\uparrow) + 2V_O^{2+} + 4e^-, \qquad (2)$$

where $V_O^{2+}$ is the oxygen vacancy [19, 20, 35]. If during the annealing of SBTN the weakly bounded Bi leave $\{Bi_2O_2\}$ layers, the donors ($V_O^{2+}$) appear and this should lead to increase of $V_O^{2+}$ amount during annealing.

The qualitative agreement of the experimental results (**Fig. 8b** and **8d**) and the results of theoretical simulations (**Fig. 3a**) speaks in favor of such a scenario. Indeed, for the same value of $t_{ca}$ = 30 min the increase of $T_{ca}$ from 750 °C to 780 °C leads to a perceptible increase in the coercive voltage and the ratio of residual charge to the maximal one along with the decrease of residual and maximal charge (see **Fig. 8b** and **8d**). It is the case demonstrated by the theory with the increase of the concentration of mobile donors from $n_1$ to $n_3$ (see **Fig. 3a**). It should be noted that the shape and characteristics of the charge-voltage loop in **Fig. 8d** is the average between the loops corresponding to $n_2$ and $n_3$, shown in **Fig. 3a**.

### 5.3. Effect of annealing time on the parameters of charge-voltage loops

At the same annealing temperature $T_{ca}$ = 750 °C the increase of $t_{ca}$ from 30 to 60 min. reduces the remanent charge under small decrease of the coercive voltage and the ratio of the remanent charge to the maximal one (see **Fig. 8b** and **8c**). These facts are in correspondence with the results of simulation (**Fig. 3a**) if the concentration of mobile donors reduces from $n_2$ to $n_1$.

It is known for the ceramics of $SrTiO_3$ and $BaTiO_3$ type [19], that during the low-temperature annealing (T < 500°C) the equilibrium with ambient oxygen atmosphere cannot be achieved within reasonable limits of its duration due to the slow kinetics of oxygen exchange through the surface. Allowing for the small value of SBTN film thickness (200 nm) one can assume that these "reasonable limits" of annealing time are restricted by several hours. At higher annealing temperatures (T > 700 °C) the rate of reaction (2) increases [19, 30, 35]. Thus, the decrease of $V_O^{2+}$ concentration may take place depending on the partial pressure of oxygen in the surrounding atmosphere, the annealing time and the interactions between oxygen defects and metal sublattices in SBTN films.

Taking this into account, we can regard results presented in **Fig. 8b** and **8c** in qualitative agreement with the results of simulation (**Fig. 3a**) for the case of reducing the concentration of mobile donors from $n_2$ to $n_1$. Considering the value of vertical displacement of the charge-voltage loop as a measure of the leakage current of the film and comparing **Fig. 8b and 8c**, one can see the reduction of this current with increasing $t_{ca}$. This corresponds to a reduction of the free carriers' amount associated with the concentration of mobile donors. Reduction of maximal polarization instead of the expected increase (see **Fig. 3a**) may be related to the decrease of dielectric contribution due to changes in the defect subsystem with $t_{ca}$ increase, not considered by the theory.

This result suggests advisability of the annealing time reduction and using a relatively fast thermal treatments for the formation of ferroelectric structures based on SBTN films. Short-time crystallization annealing is used, in particular, to obtain the perovskite phase of PZT films [2].

Full compliance between the theoretical predictions and experimental results can be achieved on the basis of advanced understanding of the specificity of formation and interaction of planar and point defects in a layered structures SBT, SBN and SBTN.

**Conclusion**

Using the continuum phenomenological Ginzburg-Landau-Devonshire-Khalatnikov theory for the spatial distribution of polarization in the film together with the continuity equation for the electron and mobile charged donors currents and the Poisson equation for electrostatic potential we modeled the charge-voltage and current-voltage characteristics for ferroelectric semiconductors with mobile charged donors.

With the temperature variation at a fixed relatively low concentration of donors the slope and the coercive voltage of nonsaturated charge-voltage loop change, as well as the magnitude and voltage position of maxima of current-voltage loop. This behavior is similar to that adopted for ferroelectrics in the low temperature vicinity of ferroelectric-paraelectric phase transition.

The increase of mobile donors concentration at constant temperature leads to the change of the charge-voltage loop shape from saturated to elliptical, which is accompanied by small non-monotonic change of the remanent polarization and increase of the coercive voltage. This is not consistent with the accepted idea of lowering the remanent polarization and coercive field with increasing carrier concentration due to lowering of the effective Curie temperature.

The changes of the shape and features of charge-voltage and the current-voltage loops of the SBTN films with different Nb content is in qualitative agreement with the results of simulations.

The change of remanent polarization and coercive voltage of the charge-voltage loops of SBTN films with the increase of annealing temperature and duration, taking into account changes in the oxygen vacancies concentration, is in qualitative agreement with the results of simulation. Deviation of maximal polarization behavior from the theoretically predicted one can be attributed to the influence of unaccounted dielectric contribution.